\documentstyle{mn}
\newif\ifAMStwofonts
\title[Prescription for Star Formation Feedback]
       {A Prescription for Star Formation Feedback:\\
        The Importance of Multiple Shell Interactions}

\author[J. Scalo and D. Chappell]
      {John Scalo and David Chappell\\
       Astronomy Department, University of Texas, Austin, TX 78712}
\date{Accepted 1999 XXXXXXX XX.
      Received 1999 January XX;
      in original form 1999 January XX}

\pagerange{\pageref{firstpage}--\pageref{lastpage}}
\pubyear{1999}

\begin{document}

\maketitle

\label{firstpage}

\begin{abstract}

The relation between the star formation rate and the kinetic energy
increase in a region containing a
large number of stellar sources is investigated as a possible prescription
for star formation feedback
in larger scale galaxy evolution simulations, and in connection with
observed scaling relations for
molecular clouds, extragalactic giant H II regions, and starburst galaxies.
The kinetic energy increase is
not simply proportional to the source input rate, but depends on the
competition between stellar power input
and dissipation due to interactions between structures formed and driven by
the star
formation.  A simple one-zone model is used to show that, in a
steady-state, the energy
increase should be proportional to the two-thirds power of the stellar
energy injection rate, with additional
factors depending on the mean density of the region and mean column density
of fragments.  The scaling
relation is tested using two-dimensional pressureless hydrodynamic
simulations of wind-driven star
formation, in which star formation occurs according to a threshold
condition on the column density
through a shell, and a large number of shells are present at any one time.
The morphology of the simulations
resembles an irregular network or web of dynamically-interacting filaments.
A set of 16 simulations in which
different parameters were varied agree remarkably with the simple
analytical prescription for the scaling relation.
Converting from wind power of massive stars to Lyman continuum luminosity
shows that the cluster wind model for
giant H II regions may still be viable.

\end{abstract}


\section{Introduction}

There is consensus that momentum and energy injection from young stars
plays an important role
in the dynamics of interstellar gas in galaxies, ranging from the evolution
of protogalaxies to the
scale of the smallest interstellar clouds.  For example, nearly all studies
of disk galaxy formation
agree that star formation (SF) feedback is a crucial ingredient in
controlling the properties of the
resulting model galaxy, and most explicitly recognize the extreme
uncertainty in the feedback
implementations now in use (e.g. Navarro \& White 1993, Cole et al. 1994,
Mihos \& Hernquist 1994, Heyl
et al. 1995, Quinn et al. 1996, Steinmetz \& Muller 1994, Steinmetz 1995,
Navarro \& Steinmetz 1997,
Samlund et al. 1997, Lia et al. 1998, Gerritsen \& Icke 1997, Sommer-Larsen
\& Vedel 1998; see however
Weinberg et al. 1997).  Basically the question is:  For a given stellar
source energy injection rate in a
localized region of a model galaxy (which is itself subject to extremely
uncertain prescriptions for the
star formation rate and initial mass function, which are not the subject of
the present paper), how much
energy should be deposited in the ambient interstellar gas?  Similar
considerations apply to models for the
formation of elliptical galaxies (e.g. Theis et al. 1992, Thomas et al.
1998) and the evolution of dwarf
galaxies (e.g. Anderson \& Burkert 1998, MacLow \& Ferrara 1999).  Feedback
from star formation may also have
dramatic effects on some aspects of structure formation in the universe on
scales
$\sim$100 kpc and even larger; see the comparison of models presented in
Ostriker \& Cen (1996), where one of
the models (``CDM \& GF" in their Table 1) includes SF feedback using the
prescription of Cen \& Ostriker
(1993).

A basic assumption of nearly all this work has been that the feedback from
SF is proportional to the
SFR, with some coefficient that may depend on the ambient gas parameters.
However the regions in which
star formation takes place are severely under-resolved in the simulations,
and the size of each simulated
region is so large that many stars or star clusters may be forming within
the region.  The point of the
present paper is that the density of SF events is so large that
interactions between shells driven by
winds, H II regions, supernova remnants, or superbubbles, will be
significant, and it is the competition
between the dissipation resulting from the interactions and the driving
effect of SF that determines
the fraction of the star formation input on subgrid scales that is
available as feedback energy at
larger scales.

A closely-related problem involves the supersonic linewidths observed in
molecular and atomic clouds in
the local Milky Way.  A large number of papers have attempted to explain
the linewidth-size scaling
relation (see Larson 1981, Elmegreen \& Falgarone 1996, Heithausen 1996,
and references therein for the
observed correlation; for recent theoretical discussions and references see
Xie 1997 and Vazquez-Semadeni,
Ballesteros-Paredes
\& Rodriguez 1997; a number of proposals are reviewed in Scalo 1987).
However the fact that the
linewidth scaling correlation occurs in clouds where self-gravity is
insignificant (Heithausen 1996) and
recent simulation results indicating that magnetic fields are incapable of
significantly retarding the
decay of the ``turbulence" (MacLow et al. 1998, Stone et al. 1998) suggest
that a stellar power source
is required.  (See, however, Zweibel 1998.)  Such sources might involve
embedded protostars (Norman \& Silk
1980) or shocks external to the clouds (Kornreich \& Scalo 1998).  In the
former case, protostellar censuses
sometimes indicate that regions have more than enough energy in winds to
balance gravity, but this alone
does not demonstrate that YSOs can power the linewidths since it is the
balance (if attainable) between
dissipation and energy production that is essential for determining the
viability of the power source.
Another important point is that a study of densities (from matching CS line
strengths to radiative transfer
calculations), linewidths, and sizes for a large sample of local regions
with massive star formation yields
no linewidth-size correlation at all (Plume et al. 1996).  If there are
``hidden" correlations of certain
combinations of variables in this data related to underlying physics, then
the appropriate combination
of variables must be suggested by a theoretical model.  The one discussed
in the present paper predicts
that the scaling relation for a stellar-powered model must jointly involve
the linewidth, size, density,
and mechanical luminosity of the stars.

The same problem arises in the interpretation of the power law
linewidth-luminosity (and linewidth-size)
relation observed for extragalactic giant H II regions (e.g. Terlevich \&
Melnick 1981; see Shields 1990
and papers in Tenorio-Tagle 1994 for general properties).  Models based on
stellar energy input (Hippelein
1986, Melnick et al. 1987) have entirely neglected the effects of
dissipation on the derived scaling
relations.  Melnick et al. (1987) used a relation $L\propto R^3\sigma^2$
(L, R, $\sigma$ are Lyman
continuum luminosity, H II region core size, and linewidth) for the
stellar-powered model, and concluded
that the data agree much better with the prediction of a simple virial
model, $L\propto R\sigma^2$ (see
Tenorio-Tagle et al. 1993 for a detailed ``cometary stirring" virial
model).  We show in the present paper
that when dissipation is included, the stellar-driven model predicts a
scaling relation similar to the virial
model. In addition, morphological evidence (e.g. Meaburn 1984, Bruhweiler
et al. 1991, Chu \& Kennicutt 1994)
strongly suggests that stllar wind-blown shells and filaments are
important.  Similar remarks apply to the correlation
of supernova and wind power sources with the energy in galactic superwinds
(Leitherer et al. 1992) and other galactic
outflows (see MacLow
\& Ferrara 1999, Martin 1999, and references therein) where dissipation may
make some of the injected energy
unavailable for outflows.

The present paper proposes a simple model for the scaling relation that
should exist between the
kinetic energy deposited (as measured by the gas velocity dispersion) and
the energy injected (as measured by
the SFR), for an equilibrium situation in which energy injection is
balanced by dissipation.  Two-dimensional
hydrodynamic simulations of a system of wind-driven shells which interact
through nonlinear advection
and form new stars by a column density threshold condition are presented
which, rather remarkably,
verify the analytical scaling relation.  The main result is that the
kinetic energy gain per unit volume
due to star formation scales as the two-thirds power of the SFR, and
inversely as the four-thirds power of
the mean density, although with some dependence on other parameters.  The
result is presented as a
subgrid prescription that can be used in larger-scale under-resolved
simulations, but also as a possible
model for the linewidth systematics of local interstellar clouds,
extragalactic giant H II regions, and
starburst galaxies.  The application to large scale ``blowout" winds from
starburst galaxies is least
warranted, however, since these flows are inherently anisotropic and depend
on the galactic gravitational
potential, effects that are not included in the present model.  Besides, we
are only concerned with the
local energy injection in the vicinity of the star-forming region, and not
with larger-scale flows which
this energy may drive.

\section[]{A Simple One-Zone Model}
Consider a model in which a large (``macroscopic") region of size R, mass
M, and mean
density $\rho$ contains a statistically significant number of internal
stellar or cluster kinetic energy sources which supply energy per unit
volume at
the rate \.N$_*$E where \.N$_*$ is the total star or cluster formation rate
per unit volume and E is the
average kinetic energy input per star or cluster.  This macroscopic region
could represent an
unresolved cell in a numerical simulation of a galaxy, or an individual
molecular cloud (idealizing the
complex structure of the ISM as though it could be partitioned into
discrete entities).  We assume that
these internal stellar energy sources drive winds that create internal
substructure with a characteristic
scale much smaller than R.  This substructure is expected to be in the form
of partial shells or
filaments.  We refer to this internal ``microscopic" substructure as if it
can be segmented into discrete
entities (``clouds" or ``fragments") whose average column density is
$\mu_{cl}$, average cross section is $\sigma$, and cloud-to-cloud velocity
dispersion is
c.  The number of clouds per unit volume, $N_{cl}$, is assumed to be
determined by
the competition between production by winds and depletion due to coalescent
collisions between the
clouds:
\begin{equation}
\frac{dN_{cl}}{dt}=\dot N_*-N^2_{cl}\ \sigma c,
\end{equation}
\def\ni{\noindent}
\ni where the second term on the RHS is the rate of cloud merging per unit
volume,
$N_{cl}/\tau_{coll}$, where $\tau_{coll}\approx(N_{cl}\sigma c)^{-1}$ is
the collisional
timescale.  The velocity dispersion of the clouds, c, which is observed as
the linewidth
for the region of size R, is controlled by competition between wind energy
injection and
collisional dissipation:
\begin{equation}
\frac{d(c^2/2)}{dt}=\frac{\dot N_*E}{\rho}-\frac{1}{2}\ \sigma c^3N_{cl},
\end{equation}
\ni where the dissipation term is just $\frac{1}{2}c^2/\tau_{coll}$.  The
division of the
energy injection rate per unit volume by the average region density $\rho$
converts to an
energy input per unit mass.  Dissipation due to drag forces could also be
included, but we choose not to
complicate the analysis.

Equilibrium self-regulation is possible for such systems,
but in general a variety of sustainable non-equilibrium behaviors are
possible, including
limit cycles (Ikeuchi et al. 1984), chaos (Scalo and Struck-Marcell 1987),
and long
periods of ``incubation" with very low star formation rates, puncuated by
bursts, if there
is an external energy source (e.g. a flux of small clouds, or repeated
shocking; Vazquez
and Scalo 1989).  Here we assume an equilibrium and test the predicted
scaling behavior.

There are two key steps in obtaining a scaling relation that depends {\it
only} on energy
balance (eq. 2):  1.  We assume that most of the mass of the system M is in
the internal
substructure.  In that case the number of clouds in the system is
$M/m_{cl}$, where $m_{cl}$ is the
average cloud mass, and the number per unit volume is then
$N_{cl}=\rho/m_{cl}$, independent of the dimension
of the macroscopic region (e.g. $M\sim\rho R^3$ in 3D, $M\sim\rho R^2$ in
2D).  2.  The mass of a cloud is the
average column density through the cloud times its mean cross section:
$m_{cl}=\mu_{cl}\sigma$.  Notice that this relation is independent of the
shape of the cloud, within
factors of order unity.  Then the product
$\sigma N_{cl}=\rho/\mu_{cl}$, independent of the cloud size.  The
equilibrium relation corresponding
to eq. 2 is then (to within factors of order unity)

\begin{equation}
c = \left(\frac{2\dot E\mu_{c1}}{\rho^2}\right)^{1/3}
\end{equation}
\ni where $\dot E=\dot N_*E$ is the total energy injection rate from the
stellar sources per unit volume.

This result should be independent of the geometric form of the internal
clouds (e.g. shells,
filaments, spheres), and the geometry of the
macroscopic region (e.g. two-dimensional or three-dimensional).  It can be
shown that if drag force,
as well as collisions, is included in
the energy equation, and if the effect of drag is to destroy clouds by
ablation (analogous to the ``leakage" in the model of Norman \& Silk's
1980) at a rate $N_{cl} c_{int}/\ell$, where $c_{int}$ is the internal
cloud sound
speed and $\ell$ is the cloud size, then the scaling relation is of the
same form as eq. (3), except
for a factor which is of order unity as long as the drag coefficient is of
order unity.

We identify $c^2/2$ with the kinetic energy increase per unit mass that
results from a stellar source input $\dot
E=\dot N_*E$.  This energy increase depends on the average column density
of the clouds and the average density of
the macroscopic region, but not on the size or shape of the clouds or the
dimension of the macroscopic region.  The
conversion of this quantity to a net energy input {\it rate} (which is what
is required in numerical simulations)
requires division by an assumed characteristic timescale for the star
formation process, as discussed below.

To compare with observations of local clouds and extragalactic giant H II
regions, let $L_{KE}$ be the mechanical
luminosity (erg/sec).  Noting that \.E/$\rho$ is the energy input rate per
unit mass, we have
\begin{equation}
\frac{\dot E}{\rho}=\frac{L_{KE}}{M}=\frac{L_{KE}}{\rho R^3}
\end{equation}

\ni where M is the mass of the region of size R.  Writing the second factor
of $\rho$ in
the denominator of eq. 3 in terms of the mean column density $\mu$ of the
macroscopic
region as $\rho=\mu/R$ gives
\begin{equation}
c =\left(\frac{L_{KE}/R^2}{\rho}\frac{\mu_{cl}}{\mu}\right)^{1/3}
\end{equation}

The unresolved cloud column densities $\mu_{cl}$ are unknown, but one might
assume that the
{\it ratio} $x\equiv\mu_{cl}/\mu$ is a constant from region to region,
assuming a kind of self-similarity,
although this assumption is extremely uncertain.  Obviously variations in
this ratio will introduce scatter in any
observational test of the scaling relation.  Assuming that $L_{KE}$ is
proportional to the total radiative
luminosity of the protostars in the macroscopic region $L_*$, and denoting
$L_*/4\pi R^2=F_*$, the radiative flux
(due to the internal power sources) from the macroscopic region, the
predicted scaling is
\begin{equation}
c\sim\left(\frac{F_* x}{\rho}\right)^{1/3}
\end{equation}

\ni or, in terms of the macroscopic column density $\mu$, size R, and
luminosity,
\begin{equation}
c\sim\left(\frac{L_*x}{\mu R}\right)^{1/3}
\end{equation}

However in sec. 4 below we show that, for massive stars, $L_{KE}$ is not
proportional to the radiative
luminosity, so these latter two relations may be of limited use.

We emphasize that these scaling relations
assume that the energy sources within the macroscopic region are numerous
enough that their outflows are
capable of at least partialy isotropizing the driven internal motions; we
do not expect the scaling relation
to apply to a macroscopic region containing, say, a single energy-injecting
star.  The large observed
luminosities of giant H II regions suggest that they must contain many
stars, but whether the number is
large enough to satisfy the assumption of the present model is uncertain;
see Chu \& Kennicutt (1994)
and Yang et al. (1996) for detailed studies of the 30 Dor region of the LMC
and the M33 H II region NGC
604, respectively.

The equilibrium scaling relation given by eq. 3 was derived for a simple
``one-zone"
model in which the spatial degrees of freedom were suppressed.  In effect
the derivation can be thought of
as  conceptualizing the ISM as a system composed of a very large number of
``clouds,"
integrating over a kinetic equation describing the one-point probability
distribution to derive ``cloud
fluid" equations for mean variables (e.g. Scalo and Struck-Marcell 1984),
ignoring gradient and advection
terms, and then examining the consequences of neglecting any time
dependence.  This is obviously a dangerous
procedure, even for the derivation of scaling relations.  For this reason,
in the next section we compare
the equilbrium result with a large number of two-dimensional hydrodynamic
simulations which were designed to
study more general aspects of ISM evolution (Chappell and Scalo 1999, CS).

\section[]{Simulations Of Wind-Driven Star Formation}

The simulations follow
the evolution of a system of interacting wind-driven shells which are
subject to nonlinear fluid advection and obey
global mass and momentum conservation.  A detailed presentation of the
models and discussion of the results are
given in a separate paper (Chappell \& Scalo 1999, hereinafter CS).  The
calculations solve hydrodynamical equations
describing a highly compressible fluid in which advection and the
corresponding ``ram pressure" completely dominate
the thermal pressure (Mach number very large), or, equivalently, in which
the effective adiabatic index
$\gamma$ is zero (as might approximately apply to the ISM because of the
nature of the
radiative cooling curve; see Vazquez-Semadeni, Passot \& Pouquet 1997).  In
this case there is no
energy equation to solve; the interactions of fluid elements are completely
inelastic.
Self-gravity and magnetic fields are neglected except that local
self-gravity is artificially introduced in the
form of a theshold instability criterion.  Newly-formed stars are assumed
to inject momentum as a wind with a
specified velocity.  A circularly-symmetric constant momentum  outflow
is injected locally whenever a new star forms at that site, and is assumed
to last for a time 10$^7$ yr
(see, for example, Leitherer1997).  We also allow for a delay time $\tau_d$
between the onset of
star formation and the initiation of the momentum input, treated as a
constant parameter.  Star formation is
assumed to occur at a threshold column density corresponding to the
gravitational instability criterion for
an expanding shell (see Elmegreen 1994, Comeron \& Torra 1994), assuming
that the growth rate of the fastest-growing
mode is a constant.  The linear perturbation analysis was generalized to
include accretion and local dilatational
shell stretching (see CS), but these effects turned out not to be important
for these simulations, as can be
understood physically (see Whitworth et al. 1994).  With the growth rate of
the fastest-growing mode assumed
constant, the criterion for star formation is simply that the column
density {\it through} a filament exceed the
critical value $w_cc_{sh}/\pi G$.  In the present work we simplify even
further by assuming that $c_{sh}$ is a
constant parameter that we vary between different simulations (series ``C"
below).  Our standard model uses
$c_{sh}=1$ km s$^{-1}$, which corresponds to a critical column density of
$10^{21}$ cm$^{-2}$.

The equations describing the evolution of the system are then
\begin{equation}
\frac{\partial\rho}{\partial t}+\nabla\cdot(\rho{\bf v})=0
\end{equation}
\begin{equation}
\frac{\partial\rho{\bf v}}{\partial t}+\nabla\cdot(\rho{\bf
vv})=\sum_{x^\prime}\frac{{\bf
x}-{\bf x^\prime}}{|{\bf x}-{\bf x^\prime}|}\ \frac{p_wN_*({\bf
x}^\prime,t)}{\tau_w}\
\end{equation}
\ni where $\rho$ is the gas surface density, $p_w$ is the total momentum
input per massive
star, $N_*({\bf x}^\prime,t)$ is the number of stars per unit area
injecting momentum at position $x^\prime$, and
$\tau_w$ is the duration of the momentum injection (10$^7$ yr here). In
practice we also experimented with models
that include the source term in the continuity equation (8), accounting for
the gas lost to star formation.  However
because the timescale for this gas depletion is so large compared to the
phenomena of interest, and in order to
calculate averages and other statistical quantitiesfrom a stationary
distribution, we have omitted the mass
depletion term for the calculations reported here.

The momentum source term in eq. 9 requires some explanation because it only
symbolicallly represents the finite
difference procedure followed in the simulations.  The position in question
is {\bf x}.  The sum is over the eight
nearest neighbor cells, at positions {\bf x}$^\prime$.  The unit vector
ensures that the momentum is directed
toward the position {\bf x}.  If there is a cluster at {\bf x}$^\prime$,
then $N_*(x^\prime$) is the number of
newly-formed stars at that position, per unit area.  The number of stars
formed in that cluster is computed from
the mass in the cluster using an adopted IMF.  The cluster mass is computed
from the mass in the simulation cell
times an assumed constant star formation efficiency.  The momentum input
$p_w=m(x^\prime)v$ is calculated using a
constant assumed wind velocity, 40 km s$^{-1}$, at a distance corresponding
to a cell size (7.8 pc in the
simulations reported here), and $m(x^\prime)$ is the fraction of the mass
released by the cluster at {\bf
x}$^\prime$ that enters the cell at {\bf x}, assuming the morphology of the
wind is circular in two dimensions.
(We do not yet address the interesting question of the effect of collimated
outflows rather than spherical
winds.)  The division by $\tau_w$ signifies that this mass and momentum are
redistributed over the lifetime of the
wind, 10$^7$ yr.  The motion of the cluster at position x$^\prime$ is taken
into account when calculating
velocities.  More details of this procedure are given in CS.

The advection terms are
differenced according to a variant of a Van Leer (1977) first-order scheme.
Modifications were made to minimize anomalous anisotropic effects
associated with the numerical viscosity
in the scheme, which would otherwise introduce artificial density and
velocity fluctuations in
expanding shells.  Details are given in CS.  The boundary conditions were
doubly-periodic.  The initial conditions
consisted of a uniform density field and a Gaussian velocity field with
prescribed power
spectrum.  We examined the effects of varying the initial power spectrum
and the resolution
(128$^2$, 256$^2$, and 512$^2$).  The scales were normalized such that the
lattice spacing
was 7.8 pc, so these resolutions correspond to a total region size of 1, 2,
and 4 kpc, but we
expect the essential results to apply to smaller or larger scales if the
size and velocity scaling is
adjusted.  A series of 256$^2$ simulations with initial energy spectrum
given by
$E(k)\propto k^4 exp k^2/k^2_0$ and $k_0=4$ is presented here.  The
simulations were integrated for about 2
Gyr, long enough for initial transients to disappear and to study the
temporal evolution of
the system.  We point out that these very long integrations were made
possible by neglect of physical processes
beside advection, and by the adoption of a first-order difference scheme.
A detailed presentation is given in
CS.  Here we are only concerned with checking the scaling relation given by
eq. 3.

All the simulations evolve into a network of irregularly shaped filaments
(in two dimensions)
which cover a large range of sizes and which are the products of the
distortion of the
originally symmetric star-forming shells by interactions with other shells
and by advection
(the distinction is not clear-cut, since most of the mass ends up in the
filaments).
Sometimes filament interactions lead to nearly spherical ``clumps" which
may or may not be
dense enough to form a star.  Often an expanding shell produced by one
``star" or ``cluster"
compresses gas along the filament in which it was born, stimulating further
star formation and
sometimes resulting in groups or chains of clusters.  The overall
filamentary structure is not
dependent on the existence of the wind energy input, but is an inevitable
result of the high
compressibility brought about by the absence of pressure; in fact,
simulations without stellar
forcing develop similar structure, although of course with no input the
structure is
eventually concentrated on large scales, and the velocities monotonically
decrease with time.
The ubiquity of similar networks of filmentary structure in simulations
that involve the
inclusion of different physical processes (e.g. cooling, pressure,
self-gravity, magnetic
fields, different types of star formation ``laws"; see Bania and Lyon 1980,
Chiang and
Prendergast 1985, Chiang and Bregman 1988, Vazquez-Semadeni et al. 1994,
Passot et al. 1995) suggests,
considering the present results, that such filamentary structures are
primarily due to highly
compressible advection.

An example of the gas density field at six different times is shown in Fig.
1.  The stellar distributions were
discussed in Scalo \& Chappell (1999), where it was shown that the
simulated model can account for the
observed power-law correlation functions of young stars in a number of
star-forming regions.  The simulations do
qualitatively resemble the morphology observed in well-resolved giant H II
regions like 30 Dor (Chu \& Kennicutt 1994,
who describe the structure as a ``complex network of expanding systems"),
in the entire LMC as mapped in H I (Kim
et al 1998) and in some local molecular clouds that exhibit a ``network of
filaments" (see Mizuno et al. 1995 for
Taurus), perhaps being ``churned" by protostellar outflows (e.g. Bally et
al. 1999 for the Circinus cloud).

\begin{figure*}
\vspace*{20.0cm}
\includegraphics{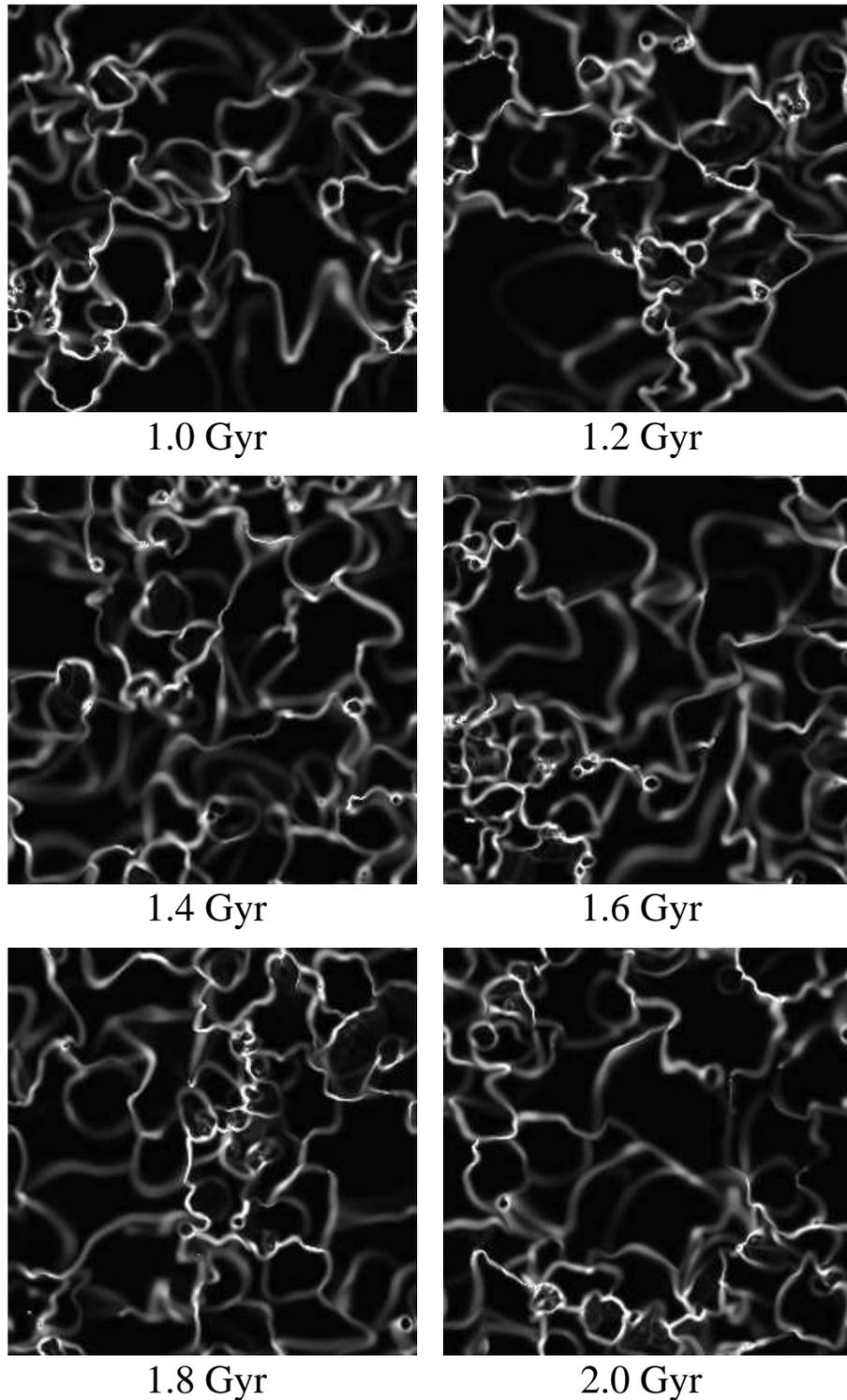}
\caption{Snapshots of the density field at six times for a typical
two-dimensional simulation.}
\end{figure*}


Out of 43 simulations we have run, we examine here a set of 16 simulations,
all with resolution
256$^2$, in which the following parameters were varied.  1.  In the series
labeled ``C" the
star formation threshold was varied by changing the assumed internal gas
velocity dispersion
in the shells, which is equivalent to varying the critical column density
for gravitational instability; the
variation covered a factor of six.  These runs effectively alter the
overall star formation rate.  2.  In
the series labeled ``D" the time delay between the onset of gravitational
instability and stellar heating
was varied between zero and 10$^7$ yr.  Increasing the time delay increases
the star formation activity and
makes the spatial and temporal behavior more coherent.  The primary reason
for this result is that the wind from
a young cluster in a filament usually disrupts the filament.  A larger time
delay increases the likelihood that
the column density of a filament, which is sweeping up ambient gas, will
reach the star formation threshold
before it is disrupted.  The net result is that more star clusters will
form, on average, per filament.  3.  In the
series labeled ``E" the assumed energy input per star formed was varied
over a factor of sixteen.  All of these
series' are different physical ways of altering the overall star level of
formation activity.

The equilibrium scaling relation eq. 3 involves the column
density $\mu_{cl}$ that corresponds to the filaments in the present
simulations.  After testing
 several approaches, the following straightforward method was adopted for
identification of
filaments.  The average density in each cell was compared to the densities
in the nearest
eight neighbors.  If the central lattice site was a maximum along at least
two of the four
possible directions, the site was considered to lie on a filament ridge.
Requiring that a
cell be a local maximum in four or more directions would locate cloud
peaks, while virtually all
sites are local maxima in at least one direction (tangent to the local
density contour).  The
local directionality of a filament was then computed by estimating the
subgrid position of the
filament ridge based on the density at the neighboring lattice sites, and
then least-square
fitting a line to the positions of neighboring lattice sites that were also
found to lie on
the filament ridge.  This method was found to give excellent results, both
by tests on
density rings with a range of widths and radii, and by visual comparison of
the filament
results with the density field of the simulations.

Figure 2 plots the time-averaged velocity dispersion for each of these
simulations as a
function of the quantity $N_*E\mu/\rho^2$.  It is seen that the simulations
conform remarkably
well with the scaling index of 1/3 predicted for the equilibrium one-zone
model.  Although
this result is encouraging, it must be remembered that the simulation
quantities plotted are
long-time averages, and we have not studied the magnitude of the predicted
fluctuations about
the mean scaling relation.  The agreement also should not be interpreted as
support for
uncritical use of equilibrium one-zone models.  However in the present case
it does appear
that the long-time averages are controlled by a balance between stellar
energy injection and
dissipative shell/filament interactions.  Another consideration is the fact
that the cloud or
filament size does not enter the scaling relation, a cancelation that
allowed us to ignore the
equation governing the number of clouds.

\begin{figure*}
\vspace*{10.5cm}
\includegraphics{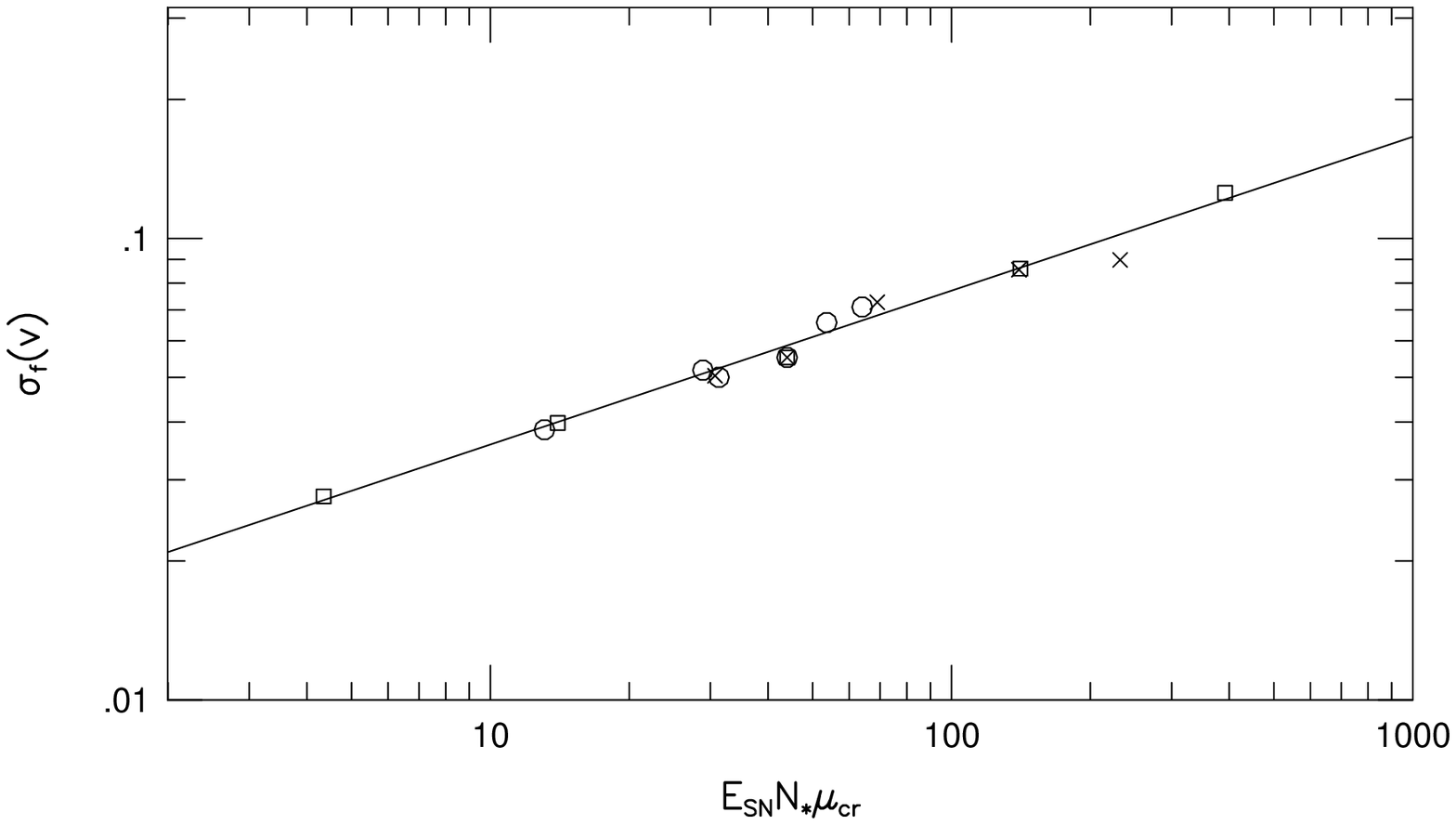}
\caption{Scaling of the filament velocity dispersion with the quantity
$\dot N_*E\mu_{cl}$ as predicted by
eq. 3. Open circles: runs C, which vary the assumed filament internal
velocity dispersion, and hence star
formation threshold.  Crosses: runs D, which vary the time delay between
filament gravitational instability
and the onset of wind energy input.  Boxes: runs E in which the assumed
kinetic energy input per massive star
was varied.  The slope of the solid line shows the scaling predicted by the
analytical model derived in the
text.}
\end{figure*}

\section[]{Discussion}
We have shown that both a simple analytical model for the balance of SF
energy injection and dissipation and
numerical simulations of wind-driven, interacting shells result in a
scaling relation between the resulting
velocity dispersion c and the energy injection rate per unit volume $\dot
N_*E$, the average column density
of ``clouds" $\mu_{cl}$, and the mean density $\rho$, given by eq. 3.  We
interpret $\epsilon\equiv c^2/2$
as the kinetic energy which results from a given SFR $\dot N_*$, assuming
that the average energy injecion
per star or star cluster E, is a constant.  Then
\begin{equation}
\epsilon\sim(\dot N_*\mu_{cl})^{2/3}/\rho^{4/3}
\end{equation}

Obviously the {\it rate} of energy feedback (which is what is needed for
galaxy simulation
prescriptions) depends on an assumed timescale, the duration of the SF
event.  If we assume that this
timescale is proportional to the collapse timescale
$\tau\sim\rho^{-1/2}$, then the density dependence of the energy injection
rate is $\rho^{-5/6}$.  However
local molecular clouds are believed to have lifetimes in excess of their
free-fall times.  Furthermore, the
simulations result in continual, ongoing star formation because of the
large number of uncorrelated
star-formation sites.  For these reasons, it might be more appropriate to
take the characteristic timescale
as a constant.  On the other hand, Elmegreen (1999) argues that a number of
physical processes should result
in a characteristic timescale that scales as $\rho^{-1/2}$.  The important
point is that the SF energy
feedback, or feedback rate, should not scale linearly with the SFR, but as
the SFR to the 2/3 power.

The additional forms of the predicated scaling relations (eqs. 5--7) should
be more useful in interpreting
the linewidths of local clouds (e.g. Plume et al. 1996) when the radiation
flux or luminosity is the
primary observed variable.  For example, in terms of the mechanical
luminosity, eq. 7 gives
$\epsilon\propto(L/\mu R)^{2/3}$, where $\mu$ is the macroscopic column
density, if the parameter x is
assumed constant.

Observations of giant H II regions give a scaling relation between velocity
dispersion
(linewidth) and Lyman continuum luminosity L of the form $c\sim L^{1/4}$ to
$L^{1/6}$ (see Hippelein 1986,
Melnick et al. 1987, 1988).  However, the observed scaling is bivariate,
with an additional dependence on
the size of the H II region, R.  The scaling relations proposed here (eqs.
3--7) are in terms of the source
kinetic energy injection rate, not the radiative luminosity.  We identify
the source kinetic energy rate with
the wind power of a massive star,
$L_w(m)\propto \dot MV_\infty^2$, where
$\dot M$ is the mass loss rate and
$V_\infty$ is the terminal wind speed, both at stellar mass m, and then
integrate over an IMF.  In order to
compare the predicted scaling relations with the observed scaling relation,
we therefore need to derive a
relation between the IMF-averaged wind power
$L_w$ and the IMF-averaged Lyman continuum luminosity $L$.

For the stars in a giant H II region, we assume a power law differential
mass spectrum of the form
$n(m)=Am^{-\gamma}$, where
$\gamma$ is a parameter.  If $\gamma>1$ (which is a very reasonable
condition), then if the Lyman continuum
luminosity varies with mass as $m^\alpha$, the IMF-averaged $L$ is
dominated by the term $Am_u^{\alpha-\gamma+1}$,
where $m_u$ is the upper mass limit.  Similarly, if the wind luminosity
varies with mass as $m^\beta$, the
IMF-averaged wind luminosity is proportional to $Am_u^{\beta-\alpha+1}$.
Eliminating $m_u$ gives $L_w\sim
L^\delta$, where
\begin{equation}
\delta=\frac{\beta-\gamma+1}{\alpha-\gamma+1}\ .
\end{equation}

The values of $\alpha$ and $\beta$ were estimated using the wind
luminosities and Lyman continuum
luminosities as a function of mass (between 23 and 87 M$_\odot$) tabulated
by Leitherer (1997, Table 6).
We evaluated the exponent $\delta$ in eq. 11 for luminosity classes V, III,
and I, although there is not
much variation in $\delta$ between these classes.  In all cases we find
that $\delta$ is significantly {\it
larger} than unity.  For LC V, $\alpha\approx3.2,\ \beta\approx2.6$, and
$\delta$ varies between 1.3 and
1.6 for IMF indices between 1 and 2.5.

For example, taking $\delta=1.5$, eq. 5 gives
\begin{equation}
L\sim\left(\frac{\rho}{\mu_{cl}/\mu}\right)^{2/3}\ R^{4/3}\ c^3
\end{equation}
\ni Earlier treatments of wind-powered velocity dispersions, which
neglected dissipation, gave $L\sim
R^3c^3$.  Melnick et al. (1987) showed that the observed bivariate
distribution does not agree with this
prediction, but gives better agreement with the virial prediction $L\sim
Rc^3$.  However the results
including dissipation (eq. 12) give a scaling that is close to the virial
scaling, although additional
parameters (mean density $\rho$ and column density parameter
$\mu_{cl}/\mu$) enter the scaling relation.
Expressed in terms of the mean column density $\mu$ (eq. 6), we find, for
$\delta=1.5,\
L\sim(R\mu/x)^{2/3}c^3 $, again close to the virial result.  Considering
that Melnick et al. (1987)
found a better, but still not exact, fit with the virial result (they found
$L\sim(Rc^3)^{0.86}$),
it appears that the wind model, including dissipation, is still viable,
although the dependence on mass
density or column density remains to be studied.  According to Kennicutt
(1984), densities are roughly
constant among giant extragalactic H II regions.  The virialization/bow
shock, or ``cometary stirring,"
model of Tenorio-Tagle et al. (1993) remains attractive; we are only
pointing out that the difference between
the scaling relations predicted by the cometary stirring model and the
stellar wind model is not nearly
as large as previously claimed.

For local molecular clouds with massive star formation, we attempted to
compare the predicted scaling (eq.
6 and 7) with the results of Plume et al. (1997), but found that the
empirical values of the luminosity,
density, and size were too uncertain to afford a meaningful comparison.
Luminosity estimates were only
available for 12 of the 150 regions studied, and a plot of linewidth versus
$L/\rho R^2$ (see eq. 5) for
these regions yielded only a very rough correlation dominated by scatter.
In addition, we have no
indication how to transform observed luminosity to kinetic energy input in
this case (they were assumed
proportional in eqs. 6 and 7).  The IMF-averaged wind-radiative luminosity
conversion derived earlier
predicts a linewidth that scales as $L^{1/2}/(\rho R^2)^{1/3}$.  When the
Plume et al. linewidths
(averages from three transitions) are plotted versus this quantity, using
their tabulated luminosities,
sizes and densities (from C$^{34}$S except in the case of GL 490 for which
only the CS density was
available), a fairly good correlation is found for most sources, but with
strong deviations for W51(OH)
(linewidth too large compared to prediction) and CRL 2591 (linewidth too
small compared to prediction).
However many of the source luminosities are so small that the energy input
cannot include stars as massive
as assumed in the IMF averages, or indicate that the number of stellar
energy sources is of order unity,
violating the assumption of our model. More observational studies aimed at
testing the prediction of the
present paper are needed.  The important point is that we showed that, in a
stellar-driven model, the scaling
relation involves several variables, all of which must be estimated.  These
predicted results should apply
just as well to clouds driven by outflows from low-mass YSO's.  We
speculate that much of the scatter found
in attempts to reconstruct the ``Larson relations" (Larson 1981) is simply
due to variations in these
additional variables.  On the other hand, an obvious weakness of the
stellar-powered model is that
Larson-type scaling is inferred for diffuse clouds without internal stellar
sources (Heithausen 1996).  This
suggests that the power source for these regions may be stellar-driven
shocks external to the clouds, along
the lines proposed by Kornreich \& Scalo (1999).  Conversion of magnetic
energy into kinetic energy by
ambipolar diffusion (Zweibel 1998) is unlikely for these clouds since the
ionized fraction should be
relatively large.

The energy input scaling derived here may also have relevance to starburst
galaxies.  Melnick et al. (1988)
showed that the scaling in small starburst ``H II galaxies" is similar to
that observed in giant H II
regions.  Also interesting is the semi-empirical correlation derived by
Leitherer et al. (1992) between the
kinetic energy of ``superwind" outflows from starburst galaxies and the
derived energy input from winds and
supernovae for a number of starburst galaxies.  The derived correlation is
{\it steeper} than linear,
compared to the less than linear prediction presented here.  However the
sizes and densities of these
starburst regions need to be included for a proper comparison with the
predicted relation.  Besides these
effects, a dependence of burst duration on metallicity could be involved,
as suggested by Leitherer
(1997).  Clearly more work is needed to compare the present model with
observations of superwinds.

We want to emphasize that the application of the present model to
superwinds and outflows from starbursting
dwarf galaxies may be inappropriate, since ``blowout" may be most effective
for an essentially single
central wind source, and involves a fight against the vertical
gravitational field of the galaxy, (e.g.
recent numerical simulations of MacLow \& Ferrara 1999).  Furthermore, as
mentioned earlier, the present
model only provides a prescription for the feedback in the local vicinity
of the star formation event, and
has nothing to say about larger scale flows driven by the feedback.

For simulations of galaxy formation and evolution the present result gives
an easy-to-implement subgrid
star formation feedback prescription.  The prescription depends on known
macroscopically-available
variables, the star formation rate and average density, except for the
presence of the term involving the
cloud column densities.

	Finally, we emphasize that the present results depend fundamentally
on the fact that the shells and shell
fragments in the model systems lose energy primarily by interactions with
ambient material, in our case other
shells.  For this situation to occur, the SFR must be large enough that the
mean separation of shell-producing
events is smaller than the size of shells when radiative losses become
important.  For example, if the shells
were due to supernova remnants expanding into a smooth medium, rather than
the cluster wind supershells assumed
here, radiative losses would begin to dominate when the shell radius is
only a few parsecs for the fiducial density
(using the analytical results given in Franco et al. 1994), leading to the
formation of a dense shell behind the
shock.  Most of the shell energy might be dissipated before an interaction
with another shell occurs, if the
average separation of SNR sources is large enough.  In this case the
one-zone self-regulation model would be
similar to previous studies of non-interacting shells (e.g. Franco et al.
1995), and a case could be made that the
energy injection rate is proportional to the first power of the SFR.
However since at least Type II supernovae,
which have a significantly larger rate than Type I events, should explode
within their parent cluster, wind-drive
bubbles should be the dominant energy-injecting process, and models for
these superbubbles remain adiabatic to
sizes much larger than the mean separation of clusters in our simulations,
so shell interactions should dominate
the dissipation.  In the case of smaller-scale winds from protostars, the
estimates given by Norman and Silk (1980)
indicate again that the shell radius at which radiative losses become
important is larger than the mean separation
of protostars, based on several estimates of the number density of
protostellar sources in molecular clouds.  For
these reasons we feel that shell interactions should dominate in a variety
of star-forming environments over a
broad range of scales, although more quantitative work is required to
establish the validity of this conclusion.  A
more detailed discussion of the implications of multiple shell interactions
is given elsewhere.

	As far as we know, the only previous studies of one-zone models
whose properties depend on the process of shell
interactions (in a galactic star formation context) are those of Norman \&
Silk (1980) and Franco \& Cox (1983)
for low-mass stars, and Franco \& Shore (1984) for massive stars.  The
Norman \& Silk model is conceptually
similar to both the analytic and simulation models presented here, since
both envision star formation as occurring
after multiple coalescence of shell fragments.  However we are unable to
deduce from their paper how the overall
velocity dispersion of the region depends on the SFR.  The Franco \& Cox
(1983) and Franco \& Shore (1984) models
are conceptually very different from the present approach.  They estimated
the number of shells that would be
required to maintain support against collapse and to fill a given volume of
star formation activity as the
criterion for self-regulation.  They then used this number to estimate the
SFR of either low-mass or massive
stars.  They were mostly concerned with the dependence of the resulting SFR
on the gas density, and it is not clear
how their results could be used to derive the scaling between kinetic
energy injection and the SFR.  A basic
difference between their model and the present model is that, in the
former, shell interactions are assumed to
inhibit  SF by preventing shells from expanding to column densities above
which stars could form, while in the
present model star formation is allowed to be induced by the shell
interactions; SF does not occur because the
shells have suffficient time to develop large column densities before
interaction.  The agreement we find between
the present one-zone models and the simulations, which allow for both types
of star formation mechanisms, suggests
that the latter, positive feedback effect, dominates.

\section*{Acknowledgments}
This work was supported by NASA Grant NAG5-3107 and a grant from Cray
Research. We thank the referee, Anthony
Whitworth, for comments and suggestions which improved the presentation.
We also thank Pepe Franco for pointing out the possibility that radiative
losses during shell expansion
might dominate dissipation by shell interactions, and for helpful
correspondence.

\bsp

\label{lastpage}

\end{document}